\documentclass[prl,aps,twocolumn,superscriptaddress,lengthcheck,showpacs,amsmath,amssymb,floatfix]{revtex4-1}

\usepackage{graphicx}
\usepackage{microtype}
\usepackage[rgb]{xcolor}
\usepackage{pdfcomment}

\hypersetup{colorlinks,linkcolor=blue,citecolor=blue,urlcolor=blue}

\newcommand{\vect}[1]{\ensuremath{\mathbf{#1}}}
\newcommand{\braket}[1]{\ensuremath{\left\langle{#1}\right\rangle}}

\begin{document}

\title{Entanglement detection in coupled particle plasmons}
\author{Javier del Pino}
\affiliation{Departamento de F{\'\i}sica Te{\'o}rica de la Materia Condensada and Condensed Matter Physics Center (IFIMAC), Universidad Aut\'onoma de Madrid, Madrid E-28049, Spain}
\affiliation{Instituto de F{\'\i}sica Fundamental, IFF-CSIC, Calle Serrano 113b, Madrid E-28006, Spain}
\author{Johannes Feist}
\affiliation{Departamento de F{\'\i}sica Te{\'o}rica de la Materia Condensada and Condensed Matter Physics Center (IFIMAC), Universidad Aut\'onoma de Madrid, Madrid E-28049, Spain}
\author{F.J. Garc{\'\i}a-Vidal}
\affiliation{Departamento de F{\'\i}sica Te{\'o}rica de la Materia Condensada and Condensed Matter Physics Center (IFIMAC), Universidad Aut\'onoma de Madrid, Madrid E-28049, Spain}
\author{Juan Jose Garc{\'\i}a-Ripoll}
\affiliation{Instituto de F{\'\i}sica Fundamental, IFF-CSIC, Calle Serrano 113b, Madrid E-28006, Spain}

\begin{abstract}
When in close contact, plasmonic resonances interact and become strongly correlated. In this work we develop a quantum mechanical model for an array of coupled particle plasmons. This model predicts that when the coupling strength between plasmons approaches or surpasses the local dissipation, a sizable amount of entanglement is stored in the collective modes of the array. We also prove that entanglement manifests itself in far-field images of the plasmonic modes, through the statistics of the quadratures of the field, in what constitutes a novel family of entanglement witnesses. Finally, we estimate the amount of entanglement, the coupling strength and the correlation properties for a system that consists of two or more coupled nanospheres of silver, showing evidence that our predictions could be tested using present-day state-of-the-art technology.
\end{abstract}

\maketitle

Surface plasmons are hybrid light-matter excitations confined at the interface between a metal and a dielectric. Due to their small mode volume and strong electromagnetic (EM) fields, surface plasmons interact very strongly with quantum optical emitters \cite{chang2006,dzsotjan10a,andersen11a,gonzaleztudela11a}, such as quantum dots\ \cite{akimov2007}, NV-centers\ \cite{kolesov2009} or inorganic \cite{gomez10a} and organic molecules \cite{bellessa04a,schwartz11a}. This, together with their broadband nature, small size and their inherent quantum properties make them a promising platform for future integrated quantum information technologies \cite{tame2013}. However, a very important problem lies in the characterization and control of those quantum properties. So far, several experiments have demonstrated that coupling photons in and out of plasmonic resonances preserves quantum features such as single-photon excitations and anti-bunching\ \cite{akimov2007}, photon-photon entanglement\ \cite{altewischer2002}, energy-time entanglement\ \cite{fasel2005} and squeezing\ \cite{huck2009}. In this work we focus on the quantum properties of the surface plasmon themselves and in particular in how many-body entanglement can be engineered using arrays of coupled plasmonic modes.

In this Letter, we present a plasmonic setup that intrinsically exhibits many-body entanglement and provide a recipe for characterizing it experimentally. Our results build on a quantum mechanical model for a 1D or a 2D array of coupled nanoparticles\ \cite{maier2002a,maier2002b,weick13} that includes the dipole-dipole interaction between particle plasmons, the losses in each nanoparticle and the possibility of injecting energy via coherent or incoherent light. Using this model we can not only study the transport of excitations through the plasmonic band, but we also demonstrate the emergence of stationary entanglement in the array at room temperature. Moreover, we argue that this entanglement can be detected by measuring fluctuations in the far-field from the light that is emitted from the plasmonic array.

We introduce three important theoretical ideas. The first one is a quantum mechanical model for the nanoparticle array that consists of an array of coupled oscillating dipoles with nearest-neighbor interaction and a local dissipation that accounts for the losses. This model results in a master equation for the density matrix associated with the plasmonic array. The second important idea is that, under very general circumstances, this density matrix will be Gaussian\ \cite{weedbrook2012} and all properties of the array can be deduced from expectation values or ``moments'' of a finite set of operators. In practice this implies a single set of exactly solvable ordinary differential equations that fully describes the evolution of the quantum surface plasmons. This technique allows us to make predictions not only on the dynamics of the dipoles (i.e., absorption and transport of energy) but also about their correlations and the resulting entanglement.

\begin{figure}[b]
\includegraphics[width=\linewidth]{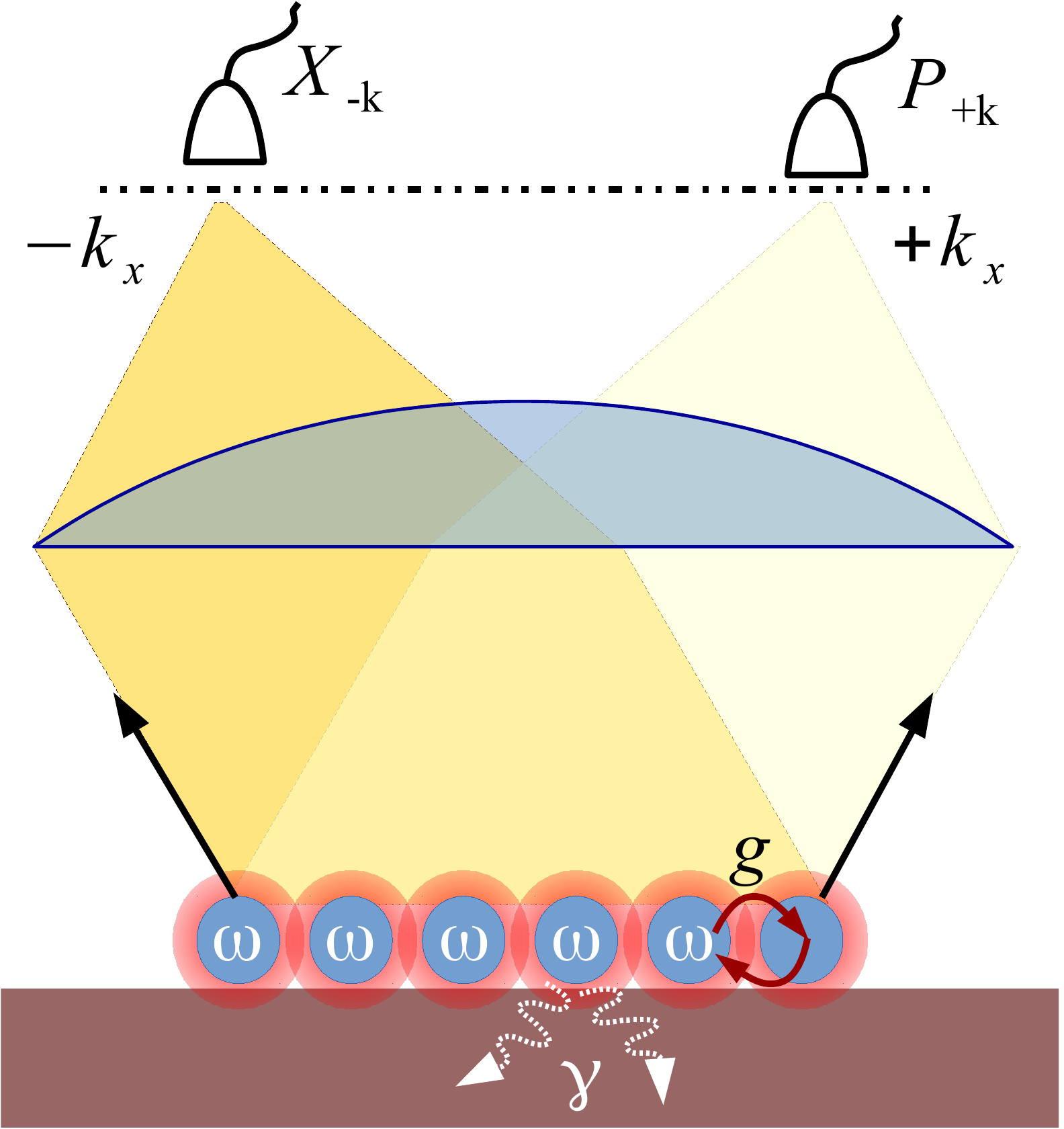}
\caption{An array of interacting nanoparticles gives rise to a set of coupled plasmonic modes. The far field emission of these modes is collected by a lens.  By correlating the properties of the light at different points in the focal plane, we get information about the multipartite entanglement.}
\label{fig:setup}
\end{figure}

The final idea in this work is a formal study of the experimental observables that can detect the presence of entanglement in the plasmonic array, the so-called \textit{entanglement witnesses}\ \cite{guehne2009,horodecki1996,lewenstein2000,terhal2000,bruss2002}. To this end, we study the plasmonic band and compute the fluctuations of the EM field in momentum space. We formally prove that the presence of squeezing in the light with opposite momenta is a signature of entanglement. From an experimental point of view, this implies that by refocusing the far-field light emitted from the structure and studying its quantum fluctuations [cf.~\autoref{fig:setup}], the amount of entanglement that is present in the plasmonic array can be quantified. This general result is valid even when the Gaussian assumption or our underlying quantum model breaks down.

%This Letter is organized as follows. We introduce an effective model for a set of coupled particle plasmons. The model results in a master equation that, in the classical limit, corresponds to the usual image of interacting dipoles. Using the %Gaussian state formalism, we derive a set of integrable equations that fully describe the static and dynamical properties of the array. A subset of these equations models the transport of plasmonic excitations when the array is driven close to %resonance. These equations relate the plasmon-plasmon interaction strength to well characterized experimental observations. We then compute the second order moments of the unperturbed array and estimate the value of the logarithmic %negativity\ \cite{adesso2007}, thus quantifying the amount of entanglement. Because the negativity cannot be measured, we derive a new observable quantity, an entanglement witness that unambiguously detects the presence of multipartite %entanglement in the plasmonic sample. We show how the value of this witness can be extracted from far-field images of the sample. The letter ends with quantitative estimates of realistic interactions and entanglement for silver nanoparticles, %together with considerations about potential experiments.

We model our coupled particle plasmons as a set of $N$ oscillating dipoles forming a linear 1D array, which interact through nearest-neighbor dipole coupling and may be subject to external driving. The Hamiltonian reads ($\hbar=1$)
\begin{equation}
H = \sum_{n=1}^{N} \frac{\omega}{2} (p_n^2 + x_n^2)+ \sum_{\langle n,m\rangle} g x_n x_m + \sum_{n=1}^N f_n(t)x_n,
\label{eq:hamiltonian}
\end{equation}
where $f_n(t)$ is a driving force, $x_n$ is the dipole moment of the particle plasmon and $p_n$ its associated canonical momentum. $g$ is the coupling strength between neighboring sites, $\braket{n,m}$, which are separated by a distance $\Lambda$. 

We introduce local dissipation by means of a master equation to describe the evolution of the quantum state or density matrix, $\rho$. This equation groups all plasmonic losses in a single parameter, $\gamma$, and reads
\begin{align}
\partial_t \rho = -\frac{i}{\hbar} [H,\rho] +
\sum_{n=1}^{N} \frac{\gamma}{2}(2a_n \rho a_n^\dagger - a_n^\dagger a_n \rho - \rho a_n^\dagger a_n),
\label{eq:model}
\end{align}
where $a_n =\frac{1}{\sqrt{2}} (x_n + i p_n)$ are the Fock operators that diagonalize each individual harmonic oscillator.

Due to the quadratic nature of the problem, we can assume that the ground state of the array is Gaussian\ \cite{weedbrook2012}, as is usually done in linear optics. This implies that the density matrix $\rho$ can be reconstructed from the expectation values, $\braket{O}:=\mathrm{tr}(O\rho)$, of the operators $O\in \{x_n,p_n, x_nx_m, p_np_m, x_np_m\}$. Moreover, the evolution equations for these ``moments'' form a closed set of first order different equations that can be exactly solved, as described in detail in section I of the Supplemental Material. Let us start with the first moment equations, which describe the dynamics of the effective dipoles $d_n=\braket{x_n}$. It is straightforward to find a set of coupled driven classical harmonic oscillators subject to friction
\begin{equation}
 \ddot{d}_{n}=-\left(\omega^{2}+\frac{\gamma^{2}}{4}\right)d_{n}-2\omega g\sum_{l}d_{l}-\gamma\dot{d}_{n}+f_{n},
  \label{eq:oscillator}
\end{equation}
where the sum over $l$ is over nearest neighbors of $n$. This is a classical model that has already been used to describe a particle plasmon array\ \cite{brongersma2000} and shows the compatibility of our master equation with earlier theoretical studies. In particular, our equations must describe the transport of excitations and absorption of energy by the plasmonic array. In fact, we can use the available experimental results to extract quantitative information about the three parameters $g, \omega$ and $\gamma$, which characterize our modeling.  

Regarding transport, let us assume a coherent driving on the first site, $f_1(t) \sim \sin(\nu t)$, and study the asymptotic state of the dipoles as a function of the distance. From this calculation we can extract a propagation length, $\xi$, defined as
\begin{equation}
\xi =\frac{\sum_{n=1}^N n \Lambda \left|\braket{x_n}\right|}{\sum_{n=1}^N \left|\braket{x_n}\right|}.
\end{equation}

\begin{figure*}[t]
\includegraphics[width=\linewidth]{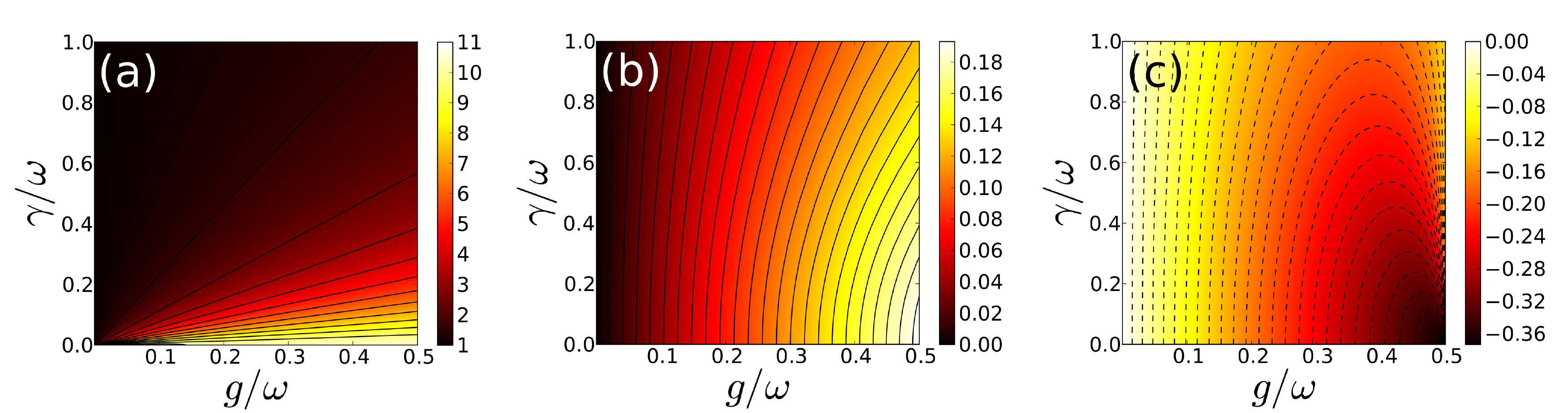}
\caption{(a) Average propagation length (in units of $\Lambda$) in the 1D chain of $N=20$ nanoparticles versus coupling strength, $g$, and local dissipation, $\gamma$. (b) Entanglement in the chain measured by the logarithmic negativity. (c) Entanglement witness in momentum space.}
\label{fig:results}
\end{figure*}

For the case of a very long chain, this propagation length would determine the exponential decay of the plasmon population, $\braket{x_n}  \sim e^{-n\Lambda/\xi}$. In \autoref{fig:results}a we show the propagation length in units of particle spacing, $\xi/\Lambda$, obtained numerically for a chain of $N=20$ oscillators, as a function of the coupling strength $g$ and plasmonic loss $\gamma$, under quasi-resonant driving ($\nu = 0.99\omega$). Dissipation leads to a finite propagation length, which grows with $g$ and diverges at the critical point $g/\omega=1/2, \gamma=0$, where the current model becomes unphysical.

%The spectroscopy of the chain under driving with different frequencies provides information about the coupling strength, $g$, and the strength of the losses, $\gamma$. In the eigenbasis of the undriven Hamiltonian (2), the driving and dissipation %force the $k$-th collective mode to absorb energy until the equilibrium condition
%\begin{equation}
%\overline{\langle X_k\rangle} \sim \frac{1}{\gamma/2 + i(\omega_k - \nu)}
%\end{equation}
%is reached. Thus, we can relate the linewidth of the resonance of the plasmonic array to the dissipation $\gamma$, and the bandwidth of the whole spectrum to the coupling strength $g$, both of which are physically observable.

While the first order moments reproduce predictions of the classical theory, the second order moments contain information about the non-classicality of the many-body particle plasmon state. In particular, the matrix of second order correlations, or covariance matrix, can also be exactly computed (see section I of the Supplemental Material) and used to quantify the amount of entanglement present in the plasmonic array. For this purpose let us eliminate the driving $f_n(t)$, whose role is merely to displace the different oscillator modes, without adding entanglement. In the absence of this driving, we focus on the second order moment for the covariance matrix
\begin{equation}
\sigma_{i,j}=\frac{1}{2}\left\{ \langle R_{i}R_{j}\rangle-\langle R_{i}\rangle\langle R_{j}\rangle\right\},
\end{equation}
where $\vect{R}^T = (x_1, \ldots, x_L, p_1, \ldots, p_L)$ is a vector that groups all positions and momenta.

Let us consider a bipartition of the plasmonic array into two subarrays, A and B. It is clear that the covariance matrix can be split into boxes that group the operators of one or the other array,
\begin{equation}
  \sigma = \left(\begin{matrix} \sigma_{AA} & \sigma_{AB} \\ \sigma_{BA} & \sigma_{BB}\end{matrix}\right),
\end{equation}
together with some off-diagonal terms, $\{\sigma_{AB},\sigma_{BA}\}$ that imply some correlation (quantum or classical) between the two arrays. In order to quantify purely quantum correlations, we compute the so called negativity\ \cite{weedbrook2012}, $E_N[\sigma;A,B]$. A value of $E_N[\sigma;A,B]$ above zero means that the plasmonic array is entangled at least with respect to this bipartition. Subsequent application of this criterion to different partitions of the array can be used to ensure true multipartite entanglement.

The results of this calculation are shown in \autoref{fig:results}b for a 1D array of 20 nanoparticles divided into two blocks of 10 consecutive particles. We plot the negativity as a function of the coupling strength $g$ and the plasmonic loss $\gamma$. As expected, entanglement grows with $g$ and becomes maximum at the critical point $g=\omega/2,\gamma=0$, where the propagation length diverges. The effect of dissipation is to decrease the entanglement, which remains sizable for moderate coupling strengths, $g \simeq \gamma$.

Unfortunately, the negativity is not an observable. It may be estimated from the full covariance matrix if a sufficiently accurate reconstruction of this matrix is available, but this is an experimentally daunting task. It would therefore be interesting to have an experimental criterion that allows the detection of entanglement in the plasmonic chain with the least number of measurements, while being robust to noise and imperfections.

For this task we suggest what is called an entanglement witness\ \cite{guehne2009,horodecki1996,lewenstein2000,terhal2000,bruss2002}. A witness is an observable $W$ such that when its expectation value $\langle W \rangle = \mathrm{Tr}(W\rho)$ becomes negative, we can positively assure that the state $\rho$ is not separable. There are several such entanglement criteria in the literature of quantum optics. One of them is the so-called Duan criterion for detecting two-mode squeezing\ \cite{duan2003}, which was later extended by Hyllus and Eisert\ \cite{hyllus2006} to include multipartite entanglement. In this work we develop a very general but simpler version of this last protocol.

\textbf{Theorem:} Let us take two vectors $\vect{u}_{1}$ and $\vect{u}_2$ which satisfy the following conditions: (i) they are normalized, $\Vert{\vect{u}_i}\Vert=1$, (ii) have the same modulus element-wise ($|u_{1,i}|=|u_{2,i}|$) and (iii) define two pairs of canonical variables,
\begin{equation}
X_k = \sum_{j=1}^L u_{k,j} x_j,\ \mathrm{and}\ P_k = \sum_{j=1}^L u_{k,j} p_j.
\end{equation}
If the two opposite quadratures are squeezed
\begin{equation}
\langle\Delta{X}_1^2\rangle + \langle\Delta{P}_2^2\rangle < 1,
\end{equation}
then the state is entangled. The demonstration of this theorem is presented in the Supplemental Material, section II.

While the conditions (i)-(ii) might seem rather artificial, they can be satisfied by the normal modes of the plasmonic array. The undriven part of Hamiltonian (1) can be diagonalized using normal modes $\{X_k,P_k\}$ (see details in section III of the Supplemental Material)
\begin{equation}
H_0= \sum_k \frac{\omega}{2}(P_k^2 + \lambda_k X_k^2),  
\end{equation}
where $k$ represents the quantized momentum, $k=\pi j/[(N+1)\Lambda]$ with $j$ running from $1$ to $N$. The magnitude $\lambda_k = 1+2(g/\omega) \cos k \Lambda$ determines the plasmonic dispersion band, $\omega_k=\omega \sqrt{\lambda_k}$.

Therefore, in the case of a 1D linear chain (corresponding to open boundary conditions) and for a very large number of nanoparticles, $\vect{u}_1$ and $\vect{u}_2$ of the theorem could be the wavefunctions associated to two eigenmodes with opposite momenta $(k,k')=(k,\pi/\Lambda-k)$, which are equal in modulus and only differ in the fact that one has alternating signs and the other does not, $u_{1,j} = (-1)^j u_{2,j}$. From a practical point of view, this means that we can detect entanglement by looking for squeezing among states with momenta $k$ and $(\pi/\Lambda-k)$. In other words, we can define our entanglement witness
\begin{equation}
W_k := \mathrm{min}\{0,\langle\Delta{X_k}^2\rangle+\langle\Delta{P}_{\pi/\Lambda-k}^2\rangle-1\}, 
\end{equation}
so that $W_k<0$ implies entanglement. For the particular case $k=0$, i.e., the extrema of the dispersion band, we can find an analytical expression for the entanglement witness (see details in section III of the Supplemental Material)
\begin{equation}
W_0=1+\frac{\frac{2g}{\omega}(\frac{2g}{\omega}-1)}{\frac{\gamma^{2}}{\omega^{2}}+4(1-\frac{2g}{\omega})}-\frac{\frac{2g}{\omega}}{\frac{\gamma^{2}}{\omega^{2}}+4(1+\frac{2g}{\omega})}.
\end{equation}

\autoref{fig:results}c presents the numerical results corresponding to $W_0$. As it shown in the plot, the growth of the witness follows the same trend as that of the negativity, hence providing the same amount of information. 

\begin{figure}
\includegraphics[width=\linewidth]{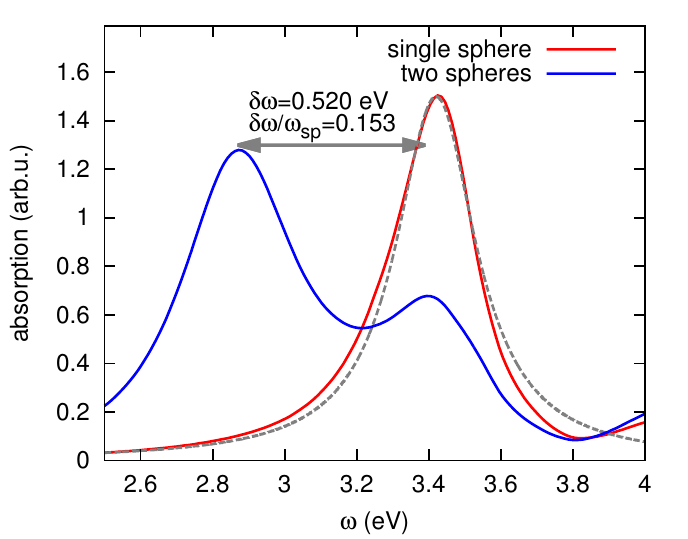}
\caption{Absorption versus frequency for a single silver nanosphere (red line) and a dimer (blue line). In these calculations the radii of the nanoparticles is set to $R=25$ nm whereas the separation between nanoparticles in the dimer case is $2$ nm. The dashed grey line represents a Lorentzian fit to the absorption spectrum of the single nanosphere that is used to estimate $\gamma$.}
\label{fig:spheres}
\end{figure}

In what follows we describe how this entanglement could be measured using present-day state-of-the-art technology. Squeezing in the plasmonic band is related to entanglement, and the same applies to far-field images of the lattice. The light emitted by the plasmons maps the quadratures in the collective variables $\{X_k,P_k\}$ onto the equivalent variables of the field propagating along directions $\pm k$. This light can be collected by a large aperture lens, so that each value of the momentum is mapped to a different point on the focal plane of the lens, as sketched in \autoref{fig:setup}. Selecting the photons with the appropriate momenta, we can perform homodyne detection\ \cite{weedbrook2012,welsch1999} to measure the quadratures and recover the value of $W_k$ mentioned above. Moreover, two important features make this a very useful protocol. The first one is that our choice of witness (i.e., momentum pairs) is not relevant, as we get similar results for other values of the momentum. This is a signature that the state is indeed many-body entangled. The second one is that while we have estimated $W_k$ using Gaussian states, the entanglement witness is valid for any physical state. In other words, measuring $W_k$ detects entanglement irrespectively of the underlying physical model.

The proposed measurements could be realized using different types of coupled plasmonic modes. One interesting possibility is provided by already existing setups with gold or silver nanoparticles\ \cite{maier2002a,maier2002b}. Earlier experiments with such nanoparticles revealed short propagation lengths, discouraging the use of such arrays for the transport of quantum information. However, in \autoref{fig:results} it can be appreciated that, while the plasmon propagation length is related to the coupling strength and local loss, there can be a non-zero amount of entanglement even when the surface plasmons do not propagate efficiently. As an example and to provide a quantitative and realistic estimation, we have calculated the EM coupling between two silver nanospheres of radii $R=25~$nm and separated by a distance of $2~$nm. As shown in \autoref{fig:spheres}, we obtain a coupling strength of around $g/\omega \approx 0.15$. By looking at the absorption spectrum for a single nanoparticle we can also extract a value for the loss coefficient, $\gamma/\omega \approx 0.08$.   These two values for $g$ and $\gamma$ are fully compatible with earlier works studying larger arrays\ \cite{harris2009}.  For this coupling and the associated plasmonic loss, we expect a measurable amount of squeezing, $12\%$ (see \autoref{fig:results}c), which would be a conclusive evidence of many-body entanglement within the plasmonic array.

Summing up, in this work we have studied a quantum model for an array of particle plasmons. The model, which can be extended to any system of interacting plasmonic resonances, not only describes the collective resonances and the transport of excitations through the system, but it also predicts the existence of many-body entanglement in the system. Using the formalism of Gaussian states and entanglement witnesses we have provided an experimental protocol to detect this entanglement and estimated the strength of the measurement outcomes for realistic setups. The entanglement witness developed in this work is quite general, as it detects entanglement in far-field images even for states that are not Gaussian, including coupled surface plasmons that do not fall within our model. Moreover, some of these ideas can be exported to other fields, such as nanophotonics, matter waves and the study of coupled resonators in superconducting circuits.

This work has been funded by the European Research Council (ERC-2011-AdG Proposal No. 290981). We also acknowledge financial support from EU FP7 project PROMISCE, CAM Research Consortium QUITEMAD (S2009-ESP-1594) and Spanish MINECO projects FIS2012-33022 and MAT2011-28581-C02-01.

\bibliographystyle{apsrev4-1}
\bibliography{nanoparticles}

\clearpage

\begin{center}
\Large{\textbf{Supplemental Material}}
\end{center}

\section*{I. Moment equations}
We develop the general framework for studying the steady state of Hamiltonians with quadratic bosonic operators. In the first place, it is convenient to define
$\boldsymbol{R}^{T}=\left(x_{1},\ldots,x_{N},p_{1},\ldots,p_{N}\right)$ and write $H$ as a quadratic form
\begin{equation}
H=\frac{1}{2}\boldsymbol{R}^{T}\boldsymbol{B}\boldsymbol{R}+\boldsymbol{F}(t)^{T}\boldsymbol{R},
\end{equation}
where $\boldsymbol{B}$ is a real, symmetric matrix and $\boldsymbol{F}(t)^{T}=\left(f_{1}(t),\ldots,f_{N}(t),0,\ldots,0\right)$ accounts for possible driving forces.

If the time evolution of a density matrix $\rho$ is given by
\begin{align}
\partial_t \rho = -i [H,\rho] +
\sum_n \frac{\gamma}{2}(2a_n \rho a_n^\dagger - a_n^\dagger a_n \rho - \rho a_n^\dagger a_n),
\label{eq:model}
\end{align}
where $a_n=\frac{1}{\sqrt{2}}(x_n+i p_n)$, we can easily show that the time evolution of the mean value of a time-independent operator $O$ is given in a compact form as 
\begin{equation}
\partial_{t}\langle O\rangle=-i\langle\left[O,H\right]\rangle+\sum_{n}\frac{\gamma}{2} \langle[a_{n}^{\dagger},O]a_{n}+a_{n}^{\dagger}\left[O,a_{n}\right]\rangle .
\label{eq:meanO}
\end{equation}
Here, we used the cyclic invariance of the trace and $\mathrm{Tr}\left\{ \dot{\rho} O \right\} =\partial_{t}\langle O \rangle$.
We apply this idea to the first and second moments of the quadratures,
$O\in\{x_{n},p_{n},x_{n}x_{m},p_{n}p_{m},x_{n}p_{m}\}$. Writing \autoref{eq:meanO} in the quadrature basis, we get
\begin{equation}\label{eq:trace}
\partial_t\langle O\rangle\!=\!-i\langle[O,H]\rangle\!+\!\sum_{nm}\frac{\Gamma_{nm}}{2}\langle [R_{m}^{\dagger},O]R_{n}\!+\!R_{m}^{\dagger}[O,R_{n}]\rangle,
\end{equation}
where $\boldsymbol{\Gamma}=\bigoplus_{n=1}^{N}\frac{\gamma}{2}\left(\begin{array}{cc}
1 & -i\\
i & 1
\end{array}\right)$
 is a matrix that contains the effective dissipation rates corresponding to operators $R_n$ and $R_m$ in this expression. The $\bigoplus$ symbol denotes the direct sum of matrices, so for a set of matrices $\{A_{n}\}$, $\bigoplus_{n}A_{n}=\mathrm{diag}(A_{1},A_{2},\ldots,A_{n})$. If we now make use of the commutation relations for quadrature operators, written in compact form as
\begin{equation}
[R_{n},R_{m}]=i\Omega_{nm}\quad\mathrm{with}\quad\boldsymbol{\Omega}=\left(\begin{array}{cc}
0 & \boldsymbol{1}_{N}\\
-\boldsymbol{1}_{N} & 0
\end{array}\right),
\end{equation}
where $\boldsymbol{1}_{N}$ denotes the $N\mathrm{-dimensional}$ identity matrix, it is straightforward to arrive to a closed set of $2N$ equations for the first moments $\langle\boldsymbol{R}\rangle$, which we write in matrix form as
\begin{equation}
\partial_{t}\langle\boldsymbol{R}\rangle=\left(\boldsymbol{W}+\boldsymbol{\Omega F}(t)\right)\langle\boldsymbol{R}\rangle
\end{equation}
where $\boldsymbol{W}=\boldsymbol{\Omega B}+\frac{i}{2}(\boldsymbol{\Omega\Gamma}+(\boldsymbol{\Gamma\Omega})^{T})$.

As discussed in the main text, the second moments give information about the nonclassical properties of the plasmonic array.  In the same spirit, we can write the following $2N\times2N$ equations for the second moments  $\langle\boldsymbol{C}\rangle=\langle\boldsymbol{R}\boldsymbol{R}^{T}\rangle$
\begin{equation}
\partial_{t}\langle\boldsymbol{C}\rangle=\boldsymbol{W}\langle\boldsymbol{C}\rangle+\langle\boldsymbol{C}\rangle\boldsymbol{W}^{T}-2\left(\boldsymbol{\Omega\Gamma\Omega}\right){}^{T}.
\end{equation}
Here, for simplicity, we have set $\boldsymbol{F}=0$ since its role is merely to displace the first moments as we discuss in the main text and has no effect on correlations.

Now, for instance, let us consider our particular case in which, $\boldsymbol{B}=\boldsymbol{A}\oplus\boldsymbol{1}_{N}$, i.e., the Hamiltonian may be written in the form
\begin{equation}
H = \frac{\omega}{2}\vect{p}^T \vect{p} + \frac{\omega}{2}\vect{x}^T A \vect{x} + \vect{f}(t)^T\vect{x},
\end{equation}
where $\vect{x}^T = (x_1,\ldots,x_N)$, $\vect{p}^T = (p_1,\ldots,p_N)$,  $\vect{f}(t)^T = (f_1(t),\ldots,f_N(t))$  
and $\boldsymbol{A}$ is a sparse matrix whose diagonal is unity, and the only other nonzero elements are those connecting nearest neighbor sites, which are given by $2g/\omega$. In this case we can write for the first moments
\begin{align}
\partial_{t}\langle x_{n}\rangle&=\omega\langle p_{n}\rangle-\frac{\gamma}{2}\langle x_{n}\rangle,\\
\partial_{t}\langle p_{n}\rangle&=-\omega\langle x_{n}\rangle-\frac{\gamma}{2}\langle p_{n}\rangle-2g\sum_{l}\langle x_{l}\rangle+f_{n},
\end{align}
where the sum over $l$ is over nearest neighbors of $n$.
We then obtain the equation for the effective dipole operator
$d_{n}=\langle x_{n}\rangle$,
\begin{equation}
\ddot{d}_{n}=-\left(\omega^{2}+\frac{\gamma^{2}}{4}\right)d_{n}-2\omega g\sum_{l}d_{l}-\gamma\dot{d}_{n}+f_{n},
\end{equation}
which describe the dynamics of a set of coupled, driven harmonic oscillators subject to friction. Additionally, the equations for the second moments read
\begin{align}
\partial_{t}\langle x_{n}x_{m}\rangle& = 
\omega\langle x_{n}p_{m}\!+\!p_{n}x_{m}\rangle-\gamma\langle x_{n}x_{m}\rangle \\ \nonumber
\partial_{t}\langle p_{n}p_{m}\rangle& = 
-\omega\langle x_{n}p_{m}+p_{n}x_{m}\rangle - \gamma\langle p_{n}p_{m}\rangle- \\ 
& -2g\sum_{l}(\langle p_{n} x_{l}\rangle+\langle p_{m}x_{l}\rangle)\\ \nonumber
\partial_{t}\langle x_{n}p_{m}\rangle& = 
-\omega\langle x_{n}x_{m}\rangle+\omega\langle p_{n}p_{m}\rangle - \\
& - \frac{\gamma}{2}\langle x_{n}p_{m}+p_{n}x_{m}\rangle - 2g\sum_{l}\langle x_{n} x_{l}\rangle\\
\partial_{t}\langle x_{n}p_{m}\rangle& = \partial_{t}\langle p_{m}x_{n}\rangle.
\end{align}
From their steady state solution in the absence of driving, $\partial_{t}\langle\boldsymbol{R}\rangle=0=\partial_{t}\langle\boldsymbol{C}\rangle$,  
we build the covariance matrix.
\begin{equation}
\sigma_{i,j}=\frac{1}{2}\left\{ \langle R_{i}R_{j}\rangle-\langle R_{i}\rangle\langle R_{j}\rangle\right\}.
\end{equation}

Then, we consider a bipartition of the plasmonic array into two
subarrays, $A$ and $B$, and compute the so-called logarithmic
negativity between 2 bipartitions $E_{N}\left[\sigma;\, A,B\right]$. This quantity can be given in terms of the absolute value of the eigenvalues of the matrix $i\boldsymbol{\Omega\sigma}$, after performing a non-physical operation known as partial transposition, as is discussed in detail in  \cite{weedbrook2012}. It can be shown that this non-physical operation is equivalent to changing the sign of the $p_i$ components of one of the subsystems.

\begin{widetext}
\section*{II. Entanglement witness: Proof}
Let us take two vectors $\vect{u}_{1}$ and $\vect{u}_2$ which satisfy these conditions: (i) they are normalized, $\Vert{\vect{u}_i}\Vert=1$, (ii) have the same modulus element-wise ($|u_{1,i}|=|u_{2,i}|$) and (iii) define two pairs of canonical variables,
$X_j = \sum_{i} u_{ji} x_i,\ \mathrm{and}\ P_j = \sum_{i} u_{ji} p_i$ with $j=1,2$.
If we now compute the fluctuations of these operators assuming that the state is  simply separable, $\rho=\bigotimes_i\rho_i$, we have
\begin{align}
\langle{(\Delta{X_1})^2}\rangle+
\langle{(\Delta{P_2})^2}\rangle =&
\langle{X_1^\dagger X_1}\rangle - \langle{X^\dagger_1}\rangle\langle{X_1}\rangle + \langle{P_2^\dagger P_2}\rangle -
\langle{P_2^\dagger}\rangle\langle{P_2}\rangle\\
=&\sum_{i,j}\left[u_{i,1}^*u_{j,1}\langle x_i x_j\rangle
+ u_{i,2}^*u_{j,2}\langle p_i p_j\rangle\right]
- \sum_{i,j} \left[u_{i,1}^* u_{j,1}\langle x_i\rangle\langle{x_j}\rangle + u_{i,2}^* u_{j,2}\langle p_i\rangle\langle{p_j}\rangle\right]\nonumber\\
\stackrel{(1)}=&\sum_i \left[|u_{i,1}|^2\langle\Delta{x}_i^2\rangle
+|u_{i2}|^2\langle\Delta{p}_i^2\rangle\right]
\stackrel{(2)}= \sum_i |u_{i,1}|^2 \left[\langle\Delta{x}_i^2\rangle+
\langle\Delta{p}_i^2\rangle\right]
\stackrel{(3)}>= \sum_i |u_{i,1}|^2 = 1.\nonumber
\end{align}
Here we have used various key ideas: In $(1)$ we use the fact that the state is separable and thus $\langle{x_ix_j}\rangle=\langle{x_i}\rangle\langle{x_j}\rangle$ whenever $i\neq j$. In $(2)$ we use the fact that both vectors have the same modulus element-wise
, $|u_{i1}|=|u_{i2}|$. Finally, in $(3)$ we use the fact that $\langle\Delta{A}^2\rangle+\langle\Delta{B}^2\rangle >= \Vert[A,B]\Vert$ and the normalization of the vectors.

This proof can be extended to treat fully all possible cases of separable states, $\rho = \sum p_i \otimes_j \rho_j^i$, which are convex linear combinations of the previous situation we have shown. In this case the only difference is that there appear additional cross-terms due to the linear combinations, but these terms can be shown to be larger than zero, thus increasing the fluctuations\ \cite{duan2003}.
\end{widetext}

\section*{III. Normal modes}
First, in this section, we are going to show how to diagonalize the effective model in the dissipation-free case. The diagonalization of the matrix $\boldsymbol{A}$ through an orthogonal transformation, $\boldsymbol{A} = \boldsymbol{U}^T \boldsymbol{D} \boldsymbol{U}$, allows us to define new canonical variables
\begin{equation}
X_k = \sum_i u_{k,i} x_i,\; P_k = \sum_i u_{k,i} p_i.
\end{equation} 
In these new quadratures, the Hamiltonian becomes
\begin{equation}
H = \sum_k \frac{1}{2}\omega\left(P_k^2 + \lambda_kX_k^2\right) + \sum_i u_{k,i}f_i(t) ,
\end{equation}
where $\lambda_k=D_{k,k}$ are the eigenvalues of the quadratic form and we introduce the effective drivings in momentum space, $\tilde{f}_k = \sum_i u_{k,i} f_i$. 
Note that in absence of driving, the normal frequencies of the problem will be
\begin{equation}
\omega_k = \omega \lambda_k,
\end{equation}
and the new Fock operators will be related to the original ones by a complicated squeezing transformation
\begin{equation}
\tilde{a}_k = \sum_i\left(\lambda_k^{1/2}u_{k,i}x_i + i \lambda_k^{-1/2} u_{k,i}p_i\right),
\end{equation}
that is the source of the entanglement of this problem.

As a particular instance of the lattice of coupled plasmons we will consider the case of a one-dimensional lattice of regularly spaced nanoparticles, with period $\Lambda$, that corresponds to the 1D open-boundary condition case of $\boldsymbol{A}$. This tridiagonal matrix is diagonalized with the orthogonal transformation
\begin{equation}\label{eq:orthogonal}
u_{i,k} = u_{k,i} = \sqrt{\frac{2}{N+1}} \sin(k_j \Lambda i),\ \ \;i,j=1\ldots N
\end{equation}
where $N$ is the total lattice size and the quasimomenta $k_j = \pi j/(N+1)\Lambda$ determine the eigenfrequencies
\begin{equation}
\lambda_k = 1 + 2(g/\omega) \cos(k_j\Lambda).
\end{equation}
Notice how for small momenta, when we reach the critical value $g=\omega/2$, we recover a linear dispersion relation of photon-like quasiparticles with diverging correlations. In practice, however, $g$ is below this limit and we obtain a band of massive excitations with a finite correlation length.

Now, as the form of \ \eqref{eq:trace} is preserved under the transformation \eqref{eq:orthogonal} we apply the ideas exposed previously to the first and second moments of the canonical variables, obtaining
\begin{align}
\label{eq:moments-1}
\partial_{t}\langle X_{k}\rangle&=\omega\langle P_{k}\rangle-\frac{\gamma}{2}\langle X_{k}\rangle \\
\partial_{t}\langle P_{k}\rangle&=-\omega\lambda_k\langle X_{k}\rangle-\frac{\gamma}{2}\langle P_{k}\rangle - \tilde{f}_k(t).\nonumber
\end{align}
For the second moments, taking $f=0$ to simplify the expressions, we get the following closed set

\begin{align}
\partial_{t}\langle X_{k}^{2}\rangle&\! = 
\omega\langle X_{k} P_{k}+P_{k}X_{k}\rangle-\gamma\langle X_{k}^{2}\rangle+\frac{\gamma}{2} \\
\partial_{t}\langle P_{k}^{2}\rangle&\! =\! 
-\omega\lambda_{k}\langle X_{k}P_{k}\! +\! P_{k}X_{k}\rangle-\gamma\langle P_{k}^{2}\rangle+\frac{\gamma}{2} \\
\partial_{t}\langle X_{k}P_{k}\rangle&\! = 
\! -\omega\lambda_{k}\langle X_{k}^{2}\rangle+\omega\langle P _{k}^{2}\rangle \! -\! \frac{\gamma}{2}\langle\!  X_{k}P_{k} \! +\!  P_{k}X_{k}\rangle \\
\partial_{t}\langle P_{k}X_{k}\rangle&\! = \partial_{t}\langle X_{k}P_{k}\rangle.
\end{align}
In this case, its steady-state solution allows us to compute the fluctuations needed when computing the proposed entanglement witness
\begin{equation}
W_k := \mathrm{min}\{0,\langle\Delta{X_k}^2\rangle+\langle\Delta{P}_{(\pi/\Lambda)-k}^2\rangle-1\}
\end{equation}
given  $(\Delta O)^{2}=\langle O^{\dagger} O \rangle-\langle O^{\dagger}\rangle\langle O\rangle$.
In the case of infinite chain length $N\gg1$, if we compute this quantity between the extrema of the band we arrive to
\begin{equation}
W_0=1+\frac{\frac{2g}{\omega}(\frac{2g}{\omega}-1)}{\frac{\gamma^{2}}{\omega^{2}}+4(1-\frac{2g}{\omega})}-\frac{\frac{2g}{\omega}}{\frac{\gamma^{2}}{\omega^{2}}+4(1+\frac{2g}{\omega})}.
\end{equation}

\end{document}